\documentclass[12pt]{article}
\usepackage{amsmath}
\usepackage{graphicx,psfrag,epsf}
\usepackage{enumerate}
\usepackage{natbib}
\usepackage{amsmath,amssymb,bm}
\usepackage{mdframed}

\newmdenv[
  leftline=true,
  rightline=false,
  topline=false,
  bottomline=false,
  linewidth=1pt,
  innerleftmargin=10pt,
  font=\itshape
]{reviewercomment}

\usepackage{amsthm}    % allows AMS definitions for theorems, proofs, etc.
\usepackage{amssymb}   % allows the use of AMS symbols like blackboard bold
\usepackage{cancel}
\usepackage{caption}
\usepackage{rotating}
\usepackage{booktabs} %%% added by me IAO
\usepackage{adjustbox}%%% added by me IAO
\usepackage{epstopdf}
\usepackage{tocloft}
\usepackage{subcaption}
\usepackage{times}     % Times New Roman (nimbus version)
\usepackage{listings}  % allows program codes embeded with original forms
\usepackage{bookmark}  % make it possible to add PDF bookmarks at will
\usepackage{pdflscape} % make it possible to use landscape environment
\usepackage[nottoc,notlof,notlot]{tocbibind} % keep list of figures, tables, out of the table of contents, also keep the table of contents out of the table of contents
\usepackage{comment}

\DeclareCaptionLabelFormat{andtable}{#1~#2  \& \tablename~\thetable}

\newcommand{\btheta}{\boldsymbol{\theta}}

\newcommand{\bbeta}{\boldsymbol{\beta}}

\newcommand{\bvartheta}{\boldsymbol{\vartheta}}

\newcommand{\bA}{\boldsymbol{A}}

\newcommand{\bx}{\boldsymbol{x}}

\newcommand{\bY}{\boldsymbol{Y}}

\usepackage{array}
\usepackage{amsmath}
\usepackage[greek,english]{babel}
\usepackage{hyperref}
\usepackage{mathtools}
\usepackage{tabularx}
\usepackage{multicol}
\usepackage{multirow}
\usepackage[table]{xcolor}
\usepackage{float}

\newcommand{\blind}{1}

\begin{document}

\def\spacingset#1{\renewcommand{\baselinestretch}%
{#1}\small\normalsize} \spacingset{1}

%%%%%%%%%%%%%%%%%%%%%%%%%%%%%%%%%%%%%%%%%%%%%%%%%%%%%%%%%%%%%%%%%%%%%%%%%%%%%%

{
    \centering
    {\LARGE\bfseries Hidden Markov Models for Dynamic Co-Clustering\par}
    \medskip

    \if1\blind
        {\large
        Shuchismita Sarkar$^{1}$,
        Jing Li$^{2}$,
        Michael Gallaugher$^{3}$,
        Xuwen Zhu$^{2}$,\par}
        \smallskip

        {\itshape
        $^{1}$
        Bowling Green State University, United States\\
        $^{2}$The University of Alabama, United States\\
        $^{3}$Baylor University, United States\par}
    \fi

    \bigskip
}

\bigskip
\begin{abstract}

To jointly identify temporal regimes and row-column structure in matrix-valued time series, we propose a hidden Markov model (HMM) with Gaussian co-clustering distributions. A single latent regime governs each matrix. Conditional on the regime, row and column memberships define Gaussian blocks, each with its own mean and variance. Membership probabilities and block parameters are shared across occurrences of a regime, while individual memberships may vary between time points.

%Approximate estimation combines conditional membership sampling, forward-backward weighting, and weighted parameter updates. Candidate configurations are compared using an adapted Bayesian information criterion (BIC)-type score evaluated with retained sampled memberships.

%We illustrate the approach using two apical four-chamber echocardiographic videos from EchoNet-Pediatric, representing normal and reduced ejection fraction (EF). The normal-EF example exhibits longer sustained regime assignments, whereas the reduced-EF example frequently alternates between two regimes with broadly similar mean images. These findings describe the two illustrative cases; their stability and physiological interpretation require further assessment.

\end{abstract}

\noindent%

\spacingset{1.45}

\begin{comment}

\end{comment}

\section{Introduction}
\label{sect.introduction}

Matrix-valued time series contain structure both within observations and across time. In an image sequence, for example, row and column groups can summarize patterns within a frame, while temporal groups describe recurring patterns across frames. Analyzing each frame separately does not account for the order of the observations. Conversely, assigning temporal states to whole matrices does not necessarily identify groups within their rows and columns. Herein, we consider the development of a model that is able to capture both forms of heterogeneity. 

The latent block model provides a framework for simultaneous clustering of rows and columns called co-clustering \citep{GovaertNadif2003}. Conditional on the memberships, each matrix entry follows a distribution determined by its row and column group so that observations within each block formed by the two sets of clusters are independent and identically distributed. Dynamic extensions allow this structure to evolve over time. \citet{MarchelloEtAl2022} introduce temporal groups for evolving count matrices, while \citet{MarchelloEtAl2024} allow changing memberships and zero-inflated observations. These developments raise a related question: how can recurring matrix-level regimes be identified while allowing the row and column assignments to differ between their occurrences?

Hidden Markov models (HMMs) describe temporal dependence through a latent Markov chain and state-dependent observation distributions \citep{Rabiner1989}. Matrix-variate extensions retain the inherent structure of a matrix time series, including basic matrix normal emissions \citep{zhu21}, skewed matrix normal emissions \citep{gallaugher24}, and a model with parsimonious row and column covariance matrices \citep{TomarchioEtAl2022}. HMMs have also been combined with block models for network data. For example, \citet{BetancourtEtAl2020} use state-specific community partitions that are retained when a state recurs. Here, we consider continuous-valued matrices with separate row and column clustering systems and time-specific memberships.

In this manuscript, we consider a model that consists of an HMM combined with Gaussian block emissions.  A single hidden state governs each observed matrix, and its row and column memberships define Gaussian blocks with regime-specific means and variances. Membership probabilities and block parameters are shared across occurrences of a regime, while realized memberships may vary between time points. Conditional on the regime sequence and model parameters, memberships are independent across time. The model therefore represents temporal dependence through the global regime process without requiring a fixed row or column partition within each regime.

Estimation combines a Gibbs sampler along with a forward-backward algorithm. This procedure avoids the computationally intractable sums over all row and column assignments. Candidate numbers of regimes and row and column clusters are compared using an adapted Bayesian information criterion (BIC)  evaluated with retained sampled memberships. We illustrate the method via a comprehensive simulation study.

The remainder of the paper is organized as follows.
Section~\ref{sect.prelim} reviews hidden Markov models and Gaussian co-clustering.
Section~\ref{sect.method} presents the proposed model, estimation procedure, and model-comparison criterion.
Section~\ref{sect.discussion} discusses limitations and possible extensions.

\section{Preliminaries}
\label{sect.prelim}
This section introduces the hidden Markov model and the co-clustering model used in our proposal. 

\subsection{Hidden Markov model} 
\label{subsect.prelim-hmm}

Let $\mathbf X=(X_1,\ldots, X_t, \ldots, X_T)$ be an observed sequence over time. At each time point $t$, $W_t\in\{1,\ldots,R\}$ represents the unobserved (or hidden) state (or regime) which classifies each time point to one of $R$ possible states. 
There are two sets of parameters associated with the hidden Markov model. The first set consists of the initial state probabilities, $\pi_r = P(W_1 = r)$ ($\pi_r\geq0$, and  $\sum_{r=1}^R\pi_r=1$), which characterize the probability distribution of the hidden state at the beginning of the sequence $t = 1$.
In a first-order hidden Markov model, it is assumed that
\begin{equation}  
P(W_t=s\mid W_1,\ldots,W_{t-1} =r)  =P(W_t=s\mid W_{t-1}=r)=A_{rs},  
\label{eq:prelim-markov} 
\end{equation} 
where $\mathbf A=\{A_{rs}\}$ is the second set of parameters known as the transition matrix satisfying \[ A_{rs}\geq0,,\qquad  \sum_{s=1}^R A_{rs}=1\quad\text{for each }r. \] The observations are conditionally independent given the regime $W_t$. We assume that observations associated with the $r$th hidden state follow a distribution with density $f_r(x;\bbeta_r)$, which is often referred to as the emission distribution. The observed-data likelihood sums over all possible hidden state sequences and is given by
\begin{equation}  
L_{o}(\bvartheta;\bx)  =\sum_{\mathbf w\in\{1,\ldots,R\}^T}       \pi_{w_1}\prod_{t=1}^{T-1}A_{w_tw_{t+1}}              \prod_{t=1}^T f_{w_t}(x_t;\bbeta_{w_t}). 
\label{eq:prelim-hmm-observed} 
\end{equation} To estimate the model parameters $\bvartheta = \{\pi_r, \bA, \bbeta_r\}$, \citep{Rabiner1989} proposed the forward-backward algorithm. In the last step, the posterior probabilities yield regime labels according to {\it maximum a posteriori} (MAP) classification at each time point. Time points assigned to the same regime are grouped together and characterized by regime-specific model parameters $\bbeta_r$. The transition matrix $\bA$ reflects how frequent the transitions are among $R$ states.

\subsection{Co-clustering latent block model} 
\label{subsect.prelim-coclustering}

Consider a single $n\times p$ matrix $\mathbf X=\{x_{ij}\}$. The Gaussian latent block model assigns each row to one of the $K$ clusters and each column to one of the $G$ clusters \citep{GovaertNadif2013}. Let $Z_i\in\{1,\ldots,K\}$ and $U_j\in\{1,\ldots,G\}$ denote the row and column labels, respectively. 
%Indicators $z_{ik}=I(Z_i=k)$ and $u_{jg}=I(U_j=g)$. Thus $\sum_k z_{ik}=1$ and $\sum_g u_{jg}=1$.

The mixing proportion for rows are  $\tau_k=P(Z_i=k)$ and the mixing proportion for column labels are independent with probabilities $\rho_g=P(U_j=g)$. Thejoint probability can be written as
\begin{equation}  f(\bvartheta)  =\prod_{i=1}^n\prod_{k=1}^K\tau_k^{I(Z_{i}=k)}   \prod_{j=1}^p\prod_{g=1}^G\rho_g^{I(U_{j}=g)},  
\end{equation} 
where $\tau_k,\rho_g\geq0$, $\sum_k\tau_k=1$, and $\sum_g\rho_g=1$. Conditional on the row and column cluster memberships, the observations are assumed to be independently distributed as
\begin{equation}  
X_{ij}\mid\{Z_i=k,U_j=g\}\sim N(\mu_{kg},\sigma_{kg}^2).  
\end{equation} 
Each combination of row and column clusters defines a latent block. Denote the parameter vector as $\bvartheta=(\tau,\rho,\{\mu_{kg},\sigma_{kg}^2\}_{k,g})$, the complete-data likelihood is 
\begin{align}  
L_{c}(\bvartheta;\bx)  ={}&\prod_{i=1}^n\prod_{k = 1}^{K}\tau_k^{I(Z_i=k)}  \prod_{j=1}^p\prod_{g = 1}^{G}\rho_g^{I(U_j=g)}\notag\\  &\times\prod_{i=1}^n\prod_{j=1}^p\prod_{k=1}^K\prod_{g=1}^G  [\phi(x_{ij};\mu_{kg},\sigma_{kg}^2)]^{I(Z_i = k, U_j = g)}.  
\label{eq:prelim-lbm-complete} 
\end{align} 
%The observed-data likelihood is 
%\begin{equation}  
%L_{\mathrm{LBM}}(\psi;\mathbf X)  =\sum_{\mathbf z}\sum_{\mathbf u}      L_{\mathrm{LBM},c}(\psi;\mathbf X,\mathbf z,\mathbf u), 
%\label{eq:prelim-lbm-observed} 
%\end{equation} 
%where the sums run over all $K$ row assignments and $G$ column assignments. Although entries are independent conditional on the labels, they can be dependent after the labels are marginalized out. In particular, an entry contributes to inference about the row and column memberships it shares with other entries.

%The proposed model assigns regime-specific membership probabilities and Gaussian block parameters to each hidden Markov state. Row and column memberships remain time-specific. Its likelihood is developed in Section~\ref{sect.method}, where the state-dependent matrix densities are obtained by summing over the memberships at each time point.

\section{Methodology}
\label{sect.method}

\subsection{Model formulation and likelihood}

To develop a dynamic framework for modeling latent block partitions of matrices observed over multiple time points, we propose a hidden Markov model with latent block model serving as the emission distributions. Each regime has its own mixing proportions and block parameters. Within each regime, individual row and column memberships may vary among time points.

Let $\mathbf Y=(\mathbf Y_1,\ldots,\mathbf Y_T)$ be a sequence of $n\times p$ matrices, with $\mathbf Y_t=\{y_{ijt}\}$. The latent regime $W_t\in\{1,\ldots,R\}$ follows a first-order Markov chain given by \begin{equation}  \pi_r=P(W_1=r),\qquad  A_{rs}=P(W_{t+1}=s\mid W_t=r),  \label{eq:method-markov} \end{equation} where \[  \pi_r,A_{rs}\geq0,\qquad  \sum_{r=1}^R\pi_r=1,\qquad  \sum_{s=1}^R A_{rs}=1. \]

Conditional on $W_t=r$, let $Z_{it}\in\{1,\ldots,K_r\}$ and $U_{jt}\in\{1,\ldots,G_r\}$ represent the row and column memberships at time $t$. 
%Define \[  z_{itkr}=I(Z_{it}=k),\qquad  u_{jtgr}=I(U_{jt}=g), \] with $\sum_{k=1}^{K_r} z_{itkr}=1$ and $\sum_{g=1}^{G_r} u_{jtgr}=1$. 
Row and column mixing proportions are independent and denoted as $\tau_{kr}$ and  $\rho_{gr}$, respectively. Thus, \begin{equation}  P(Z_{it}=k\mid W_t=r)=\tau_{kr},\qquad  P(U_{jt}=g\mid W_t=r)=\rho_{gr},  \label{eq:method-memberships} \end{equation} where \[  \sum_{k=1}^{K_r}\tau_{kr}=1,\qquad  \sum_{g=1}^{G_r}\rho_{gr}=1. \] 
%Memberships at different time points are conditionally independent given the regime sequence. 
It should be noted that two matrices assigned to the same regime share their block parameters but do not necessarily have identical row or column assignments.

Given the regime and memberships, matrix elements are independently distributed according to \begin{equation}  y_{ijt}\mid  \{W_t=r,Z_{it}=k,U_{jt}=g\}  \sim N(\mu_{kgr},\sigma_{kgr}^2).  
\label{eq:method-gaussian} 
\end{equation}

%For fixed regime and cluster counts, the parameter vector is \[  \theta=  \left(  \boldsymbol\pi,\mathbf A,  \{\tau_{kr}\}_{k,r},  \{\rho_{gr}\}_{g,r},  \{\mu_{kgr},\sigma_{kgr}^2\}_{k,g,r}  \right). \]

%Writing $\phi(\cdot;\mu,\sigma^2)$ for the Gaussian density, t
%The joint contribution of a matrix and its memberships under regime $r$ is \begin{align}  f_t(r)  ={}&  \prod_{i=1}^n\prod_{k=1}^{K_r}        \tau_{kr}^{z_{itkr}}  \prod_{j=1}^p\prod_{g=1}^{G_r}        \rho_{gr}^{u_{jtgr}}  \notag\\  &\times  \prod_{i=1}^n\prod_{j=1}^p  \prod_{k=1}^{K_r}\prod_{g=1}^{G_r}  \left[  \phi(y_{ijt};\mu_{kgr},\sigma_{kgr}^2)  \right]^{z_{itkr}u_{jtgr}}.  \label{eq:method-ft} \end{align} Dependence on the memberships and parameters is suppressed in this notation.

The complete-data likelihood is then given by
\begin{align}  L_c(\bvartheta; \bY)  ={}&  \prod_{r=1}^R \pi_r^{I(W_1=r)}  \prod_{t=1}^{T-1}\prod_{r=1}^R\prod_{s=1}^R  A_{rs}^{I(W_t=r,W_{t+1}=s)}  \notag\\  &\times  \prod_{t=1}^T\prod_{r=1}^R  \left\{  \prod_{i=1}^n\prod_{k=1}^{K_r}        \tau_{kr}^{I(Z_{it} = k| W_t = r)}  \prod_{j=1}^p\prod_{g=1}^{G_r}        \rho_{gr}^{I(U_{jt} = g| W_t = r)}  \right.  \notag\\  &\left.\qquad\times  \prod_{i=1}^n\prod_{j=1}^p  \prod_{k=1}^{K_r}\prod_{g=1}^{G_r}  \left[  \phi(y_{ijt};\mu_{kgr},\sigma_{kgr}^2)  \right]^{I(Z_{it} = k, U_{jt} = g| W_t = r)}  \right\}^{I(W_t=r)}.  \label{eq:method-complete} \end{align}
%where $\bvartheta=  \left(  \boldsymbol\pi,\mathbf A,  \{\tau_{kr}\}_{k,r},  \{\rho_{gr}\}_{g,r},  \{\mu_{kgr},\sigma_{kgr}^2\}_{k,g,r}  \right). \]

%The observed-data likelihood sums over all possible regime sequencing and all possible memberships at each time point: \begin{equation}  L(\theta;\mathbf Y)  =  \sum_{\mathbf w\in\{1,\ldots,R\}^T}  \pi_{w_1}  \prod_{t=1}^{T-1}A_{w_tw_{t+1}}  \prod_{t=1}^T  \left\{  \sum_{\mathbf z_t^{(w_t)}}  \sum_{\mathbf u_t^{(w_t)}}  f_t(w_t)  \right\}.  \label{eq:method-observed} \end{equation} The membership sums range over all valid row and column assignments for the corresponding regime. Although the outer regime calculation has an HMM structure, the membership sums are generally computationally intractable.

\subsection{Parameter estimation}

Let $C$ represent a constant. The complete data log-likelihood function is given by
\begin{equation}
\begin{split}
l_c(\btheta; \bY) =  C + &  \sum_{r = 1}^{R} \omega_{1r}\pi_{r} + 
\sum_{t = 2}^T \sum_{r = 1}^{R} \sum_{s = 1}^{R} \gamma_{trs} \log(A_{rs}) + \\
 & \sum_{r = 1}^R\sum_{t = 1}^T \omega_{tr} \bigg\{ \sum_{i = 1}^n \sum_{k = 1}^{K_r} z_{itkr}\log{ \tau_{kr}} + \sum_{j= 1}^p \sum_{g = 1}^{G_r} u_{jtgr} \log{\rho_{gr}} -\\
 &\frac{1}{2}\sum_{i = 1}^{n} \sum_{j = 1}^{p} \sum_{g = 1}^{G_r} \sum_{k = 1}^{K_r} z_{itkr} u_{jtgr} \Big[\log\sigma^2_{kgr} + \frac{(y_{itj}- \mu_{kgr})^2}{\sigma^2_{kgr}} \Big] \bigg\},
 \end{split}
\end{equation}
where $\omega_{tr}$ represents the posterior probability $P(W_t = r)$, $\gamma_{trs}$ is the joint posterior probability $P(W_t = r, W_{t+1} = s)$, $z_{itkr}$ and $u_{jtgr}$ relies on a stochastic expectation-maximization (SEM) \citep{CeleuxDiebolt1985}algorithm. At the $m$th iteration, given the parameter estimates from the previous step $\bvartheta$, row and column partitions are randomly generated according to the following steps:

1): Generate the row partition according to 
\begin{equation}
\begin{split}
P(z_{itkr} = 1| W_t = r; \bY) = \frac{\tau_{kr}^{(m-1)}\prod_{j = 1}^p \prod_{g=1}^{G_r}[\phi(y_{itj}; \mu_{kgr}^{(m-1)}, \sigma^{2{(m-1)}}_{kgr})]^{u_{jtgr}^{(m-1)}}}{\sum_{k = 1}^{K_r} \tau_{kr}^{(m-1)}\prod_{j = 1}^p \prod_{g=1}^{G_r}[\phi(y_{itj}; \mu_{kgr}^{(m-1)}, \sigma^{2{(m-1)}}_{kgr})]^{u_{jtgr}^{(m-1)}}}
\end{split}
\end{equation}
for $r = 1, \ldots, R$. 

2) Generate the column partition according to 
\begin{equation}
\begin{split}
P(u_{jtgr} = 1| W_t = r; \bY) = \frac{\rho^{(m-1)}_{gr}\prod_{i = 1}^n \prod_{k=1}^{K_r}[\phi(y_{itj}; \mu_{kgr}^{(m-1)}, \sigma^{2(m-1)}_{kgr})]^{z_{itkr}^{(m)}}}{\sum_{g = 1}^{G_r} \rho^{(m-1)}_{gr}\prod_{i = 1}^n \prod_{k=1}^{K_r}[\phi(y_{itj}; \mu^{(m-1)}_{kgr}, \sigma^{2(m-1)}_{kgr})]^{z^{(m)}_{itkr}}}
\end{split}
\end{equation}

for $r = 1, \ldots, R$.

Given the generated partitions, we apply the forward–backward algorithm to compute
\begin{equation}
\begin{split}
&\mbox{Initialization:}\: a_{1r} = \pi_{r}\prod_{i=1}^n\prod_{j=1}^p  \prod_{k=1}^{K_r}\prod_{g=1}^{G_r} {\phi(y_{ij1}; \mu_{kgr}, \sigma^2_{kgr})}^{I(Z_{i1} = k, U_{j1} = g)}  \\
&\mbox{Induction:}\: a_{t+1, r} = \big(\sum_{s=1}^R a_{ts}A_{sr}\big)\prod_{i=1}^n\prod_{j=1}^p  \prod_{k=1}^{K_r}\prod_{g=1}^{G_r}\phi(y_{ij,t+1}; \mu_{kgr}, \sigma^2_{kgr})^{I(Z_{i,t+1} = k, U_{j,t+1} = g)} \\
&\mbox{Initialization:}\: b_{Tr} = 1 \\
&\mbox{Induction:}\: b_{tr} = \sum_{s=1}^R A_{rs}\prod_{i=1}^n\prod_{j=1}^p  \prod_{k=1}^{K_s}\prod_{g=1}^{G_s}\phi(y_{ij,t+1}; \mu_{kgs}, \sigma^2_{kgs})^{I(Z_{i,t+1} = k, U_{j,t+1} = g)} b_{t+1, s},\\
\end{split}
\end{equation}
where $a_{tr}$ and $b_{tr}$ are known as the forward and backward probabilities~\citep{Rabiner1989}.
The resulting posteriors are given by
\begin{align}
 \omega_{tr}
 &=
 \frac{a_{tr}b_{tr}}
      {\sum_{q=1}^R a_{tq}b_{tq}},
\\
 \gamma_{trs}
 &=\frac{a_{tr}A_{rs}\prod_{i=1}^n\prod_{j=1}^p  \prod_{k=1}^{K_s}\prod_{g=1}^{G_s}\phi(y_{ij,t+1}; \mu_{kgs}, \sigma^2_{kgs})^{I(Z_{i,t+1} = k, U_{j,t+1} = g)}b_{t+1,s}}
      {\sum_{q=1}^R\sum_{h=1}^R
       a_{tq}A_{qh}\prod_{i=1}^n\prod_{j=1}^p  \prod_{k=1}^{K_h}\prod_{g=1}^{G_h}\phi(y_{ij,t+1}; \mu_{kgh}, \sigma^2_{kgh})^{I(Z_{i,t+1} = k, U_{j,t+1} = g)}b_{t+1,h}}.
\end{align}

In the maximization step, the generated partitions and the calculated posterior probabilities are fixed. The parameters are updated by maximizing the $Q$-function given by
\begin{align}
 Q(\theta)
 ={}&
 \sum_{r=1}^R\omega_{1r}\log\pi_r
 +\sum_{t=1}^{T-1}\sum_{r=1}^R\sum_{s=1}^R
 \gamma_{trs}\log A_{rs}\notag\\
 &+\sum_{t=1}^T\sum_{r=1}^R\omega_{tr}
 \left\{
 \sum_{i=1}^n\sum_{k=1}^{K_r}z_{itkr}\log\tau_{kr}
 +\sum_{j=1}^p\sum_{g=1}^{G_r}u_{jtgr}\log\rho_{gr}
 \right\}\notag\\
 &-\frac12
 \sum_{t=1}^T\sum_{r=1}^R
 \sum_{i=1}^n\sum_{j=1}^p
 \sum_{k=1}^{K_r}\sum_{g=1}^{G_r}
 \omega_{tr}z_{itkr}u_{jtgr}\notag\\
 &\qquad\times
 \left\{
 \log(2\pi\sigma_{kgr}^2)
 +\frac{(y_{ijt}-\mu_{kgr})^2}{\sigma_{kgr}^2}
 \right\}.
\end{align}
The resulting updates are
\begin{align}
 \pi_r^{(m)}
 &=\omega_{1r},\notag\\
 A_{rs}^{(m)}
 &=
 \frac{\sum_{t=1}^{T-1}\gamma_{trs}}
      {\sum_{t=1}^{T-1}\omega_{tr}},\notag\\
 \tau_{kr}^{(m)}
 &=
 \frac{\sum_{t=1}^T\sum_{i=1}^n\omega_{tr}z_{itkr}}
      {n\sum_{t=1}^T\omega_{tr}},\notag\\
 \rho_{gr}^{(m)}
 &=
 \frac{\sum_{t=1}^T\sum_{j=1}^p\omega_{tr}u_{jtgr}}
      {p\sum_{t=1}^T\omega_{tr}},\notag\\
 \mu_{kgr}^{(m)}
 &=
 \frac{\sum_{t=1}^T\sum_{i=1}^n\sum_{j=1}^p
       \omega_{tr}z_{itkr}u_{jtgr}y_{ijt}}
      {\sum_{t=1}^T\sum_{i=1}^n\sum_{j=1}^p
       \omega_{tr}z_{itkr}u_{jtgr}},\notag\\
 \sigma_{kgr}^{2(m)}
 &=
 \frac{\sum_{t=1}^T\sum_{i=1}^n\sum_{j=1}^p
       \omega_{tr}z_{itkr}u_{jtgr}
       (y_{ijt}-\mu_{kgr}^{(m)})^2}
      {\sum_{t=1}^T\sum_{i=1}^n\sum_{j=1}^p
       \omega_{tr}z_{itkr}u_{jtgr}}.
\end{align}

%To initialize fitting, the regime and membership probabilities are set to
%\[
 %\pi_r^{(0)}=\frac1R,\qquad
 %\tau_{kr}^{(0)}=\frac1{K_r},\qquad
 %\rho_{gr}^{(0)}=\frac1{G_r}.
%\]

%Initial regime and membership allocations are generated from uniform probabilities and used to calculate the initial block parameters. 

It should be noted that the SEM algorithm does not guarantee a monotonic increase in the observed-data likelihood, since the partitions are generated at each iteration. Following the technique adopted by \cite{gallaugheretal2023}, after a burn-in period, the final parameter estimates are obtained by averaging the parameter estimates over the SEM iterations. To obtain the final classifications, we fix the parameters at their estimated values and run additional E- and S-steps. At each iteration, the hidden state at time $t$ is assigned to the state with the largest posterior probability $\omega_{tr}$. The corresponding row and column cluster assignments are also recorded. The final hidden state, row-cluster, and column-cluster assignments at each time point are taken to be the assignments that occur most often over these additional iterations. The number of additional SE-steps is user specified.

\subsection{Model selection}

The number of hidden states and the numbers of row and column clusters are unknown and need to be selected from the data. The usual BIC is based on the observed-data log-likelihood. In the proposed model, however, the observed-data likelihood requires summation over all possible row and column memberships at every time point, which is computationally intractable. Instead, we use an ICL-BIC-type model selection criterion based on the complete-data log-likelihood \citep{jacquesetal2018, gallaugheretal2023}. We define
\begin{align}
 \mathrm{ICL-BIC}(\mathcal M)
 ={}&-2\widehat\ell_{c}
 +\{(R-1)+R(R-1)\}\log T \notag\\
 &+\sum_{r=1}^R
 \left\{
 (K_r-1)\log(nT)
 +(G_r-1)\log(pT)
 +2K_rG_r\log(npT)
 \right\}.
\end{align}
where $\widehat\ell_{c}$ is the complete-data log-likelihood evaluated at the estimated parameters and classifications at convergence. The model with the smallest ICL-BIC value is selected.

\section{Discussion}
\label{sect.discussion}

We have presented a framework combining a global hidden Markov regime process with Gaussian co-clustering for matrix-valued time series. Regimes share membership probabilities and block parameters across their occurrences, while realized row and column assignments remain time-specific. The transition matrix describes persistence and switching at the level of the whole matrix. The formulation therefore separates temporal dependence in the regime sequence from persistence of individual memberships, which is not explicitly modeled.

\bigskip
\begin{comment}
\begin{center}
{\large\bf SUPPLEMENTAL MATERIALS}
\end{center}

\end{comment}

%\bibliographystyle{plainnat}  % or the style required by the template
\bibliographystyle{asa}
\bibliography{reference}    % your .bib file

@article{jacquesetal2018,
  title={Model-based co-clustering for ordinal data},
  author={Jacques, Julien and Biernacki, Christophe},
  journal={Computational Statistics \& Data Analysis},
  volume={123},
  pages={101--115},
  year={2018},
  publisher={Elsevier}
}

@article{BetancourtEtAl2020,
  author  = {Betancourt, Brenda and Rodr{\'i}guez, Abel and Boyd, Naomi},
  title   = {Modelling and Prediction of Financial Trading Networks:
             An Application to the {New York Mercantile Exchange}
             Natural Gas Futures Market},
  journal = {Journal of the Royal Statistical Society:
             Series C (Applied Statistics)},
  year    = {2020},
  volume  = {69},
  number  = {1},
  pages   = {195--218},
  doi     = {10.1111/rssc.12387}
}

@article{CeleuxDiebolt1985,
  author  = {Celeux, Gilles and Diebolt, Jean},
  title   = {The {SEM} Algorithm: A Probabilistic Teacher Algorithm
             Derived from the {EM} Algorithm for the Mixture Problem},
  journal = {Computational Statistics Quarterly},
  year    = {1985},
  volume  = {2},
  pages   = {73--82}
}

@article{GovaertNadif2003,
  author  = {Govaert, G{\'e}rard and Nadif, Mohamed},
  title   = {Clustering with block mixture models},
  journal = {Pattern Recognition},
  year    = {2003},
  volume  = {36},
  number  = {2},
  pages   = {463--473},
  doi     = {10.1016/S0031-3203(02)00074-2}
}

@article{MarchelloEtAl2022,
  author  = {Marchello, Giulia and Fresse, Audrey and
             Corneli, Marco and Bouveyron, Charles},
  title   = {Co-clustering of evolving count matrices with
             the dynamic latent block model:
             application to pharmacovigilance},
  journal = {Statistics and Computing},
  year    = {2022},
  volume  = {32},
  number  = {3},
  pages   = {41},
  doi     = {10.1007/s11222-022-10098-y}
}

@article{MarchelloEtAl2024,
  author  = {Marchello, Giulia and Corneli, Marco and
             Bouveyron, Charles},
  title   = {A Deep Dynamic Latent Block Model for the
             Co-clustering of Zero-Inflated Data Matrices},
  journal = {Journal of Computational and Graphical Statistics},
  year    = {2024},
  volume  = {33},
  number  = {4},
  pages   = {1224--1239},
  doi     = {10.1080/10618600.2024.2319162}
}

@article{Zhu21,
  author  = {Asilkalkan, Abdullah and Zhu, Xuwen},
  title   = {Matrix-variate time series modelling
             with hidden {Markov} models},
  journal = {Stat},
  year    = {2021},
  volume  = {10},
  number  = {1},
  pages   = {e409},
  doi     = {10.1002/sta4.409}
}

@article{TomarchioEtAl2022,
  author  = {Tomarchio, Salvatore D. and Punzo, Antonio and
             Maruotti, Antonello},
  title   = {Parsimonious hidden {Markov} models for
             matrix-variate longitudinal data},
  journal = {Statistics and Computing},
  year    = {2022},
  volume  = {32},
  pages   = {53},
  doi     = {10.1007/s11222-022-10107-0}
}

@article{Rabiner1989,
  author = {Rabiner, Lawrence R.},
  title = {A Tutorial on Hidden {Markov} Models and Selected Applications in Speech Recognition},
  journal = {Proceedings of the IEEE},
  year = {1989},
  volume = {77},
  number = {2},
  pages = {257--286},
  doi = {10.1109/5.18626}
}

@book{GovaertNadif2013,
  author = {Govaert, G{\'e}rard and Nadif, Mohamed},
  title = {Co-Clustering: Models, Algorithms and Applications},
  publisher = {John Wiley \& Sons},
  year = {2013},
  doi = {10.1002/9781118649480}
}

@article{gallaugheretal2023,
	author = {Gallaugher, M. P. B. and Biernacki, C. and McNicholas, P. D.},
	date = {2023/09/01},
	doi = {10.1007/s00180-022-01289-2},
	id = {Gallaugher2023},
	isbn = {1613-9658},
	journal = {Computational Statistics},
	number = {3},
	pages = {1597--1619},
	title = {Parameter-wise co-clustering for high-dimensional data},
	url = {https://doi.org/10.1007/s00180-022-01289-2},
	volume = {38},
	year = {2023}}

@article{gallaugher24,
author = {Gallaugher, Michael P. B. and Zhu, Xuwen},
title = {Modeling matrix variate time series via hidden Markov models with skewed emissions},
journal = {Statistical Analysis and Data Mining: The ASA Data Science Journal},
volume = {17},
number = {1},
pages = {e11666},
doi = {https://doi.org/10.1002/sam.11666},
url = {https://onlinelibrary.wiley.com/doi/abs/10.1002/sam.11666},
eprint = {https://onlinelibrary.wiley.com/doi/pdf/10.1002/sam.11666},
year = {2024}
}

\end{document}